\newlength{\extralineskip}

\documentstyle[draft]{article}
\makeatletter
\newdimen\normalarrayskip              % skip between lines
\newdimen\minarrayskip                 % minimal skip between lines
\normalarrayskip\baselineskip
\minarrayskip\jot
\newif\ifold             \oldtrue            \def\new{\oldfalse}
\def\arraymode{\ifold\relax\else\displaystyle\fi} % mode of array enrties
\def\eqnumphantom{\phantom{(\theequation)}}     % right phantom in eqnarray
\def\@arrayskip{\ifold\baselineskip\z@\lineskip\z@
     \else
     \baselineskip\minarrayskip\lineskip2\minarrayskip\fi}
\def\@arrayclassz{\ifcase \@lastchclass \@acolampacol \or
\@ampacol \or \or \or \@addamp \or
   \@acolampacol \or \@firstampfalse \@acol \fi
\edef\@preamble{\@preamble
  \ifcase \@chnum
     \hfil$\relax\arraymode\@sharp$\hfil
     \or $\relax\arraymode\@sharp$\hfil
     \or \hfil$\relax\arraymode\@sharp$\fi}}
\def\@array[#1]#2{\setbox\@arstrutbox=\hbox{\vrule
     height\arraystretch \ht\strutbox
     depth\arraystretch \dp\strutbox
     width\z@}\@mkpream{#2}\edef\@preamble{\halign \noexpand\@halignto
\bgroup \tabskip\z@ \@arstrut \@preamble \tabskip\z@ \cr}%
\let\@startpbox\@@startpbox \let\@endpbox\@@endpbox
  \if #1t\vtop \else \if#1b\vbox \else \vcenter \fi\fi
  \bgroup \let\par\relax
  \let\@sharp##\let\protect\relax
  \@arrayskip\@preamble}
\def\eqnarray{\stepcounter{equation}%
              \let\@currentlabel=\theequation
              \global\@eqnswtrue
              \global\@eqcnt\z@
              \tabskip\@centering
              \let\\=\@eqncr
              $$%
 \halign to \displaywidth\bgroup
    \eqnumphantom\@eqnsel\hskip\@centering
    $\displaystyle \tabskip\z@ {##}$%
    &\global\@eqcnt\@ne \hskip 2\arraycolsep
         $\displaystyle\arraymode{##}$\hfil
    &\global\@eqcnt\tw@ \hskip 2\arraycolsep
         $\displaystyle\tabskip\z@{##}$\hfil
         \tabskip\@centering
    &{##}\tabskip\z@\cr}
\makeatother
\def\tr#1{{\rm tr}\kern-3pt\left[#1\right]}

\def\nn{\nonumber}

\def\beq{\begin{equation}}
\def\eeq{\end{equation}}
\def\be{\beq\new\begin{array}{c}}
\def\ee{\end{array}\eeq}

\renewcommand{\theequation}{\thesection.\arabic{equation}}

\begin{document}

\begin{titlepage}
\setcounter{footnote}0
\begin{center}

%\today
hep-th/9511126
\phantom . \hfill ITEP-M5/95 \\
\phantom . \hfill OU-HET-227 \\
%{\it  preliminary draft}
\vspace{0.3in}

{\Large{INTEGRABILITY and SEIBERG-WITTEN THEORY}}

{\Large{Curves and Periods}}
\\[.2in]
{\it H.Itoyama\footnote{Departement of Physics, Graduate School of Science,
Osaka University, Toyonaka, Osaka 560, Japan.
E-mail address: itoyama@funpth.phys.sci.osaka-u.ac.jp} and
A.Morozov\footnote{117259, ITEP, Moscow,  Russia.
E-mail address: morozov@vxitep.itep.ru}}
\\

\end{center}
\bigskip
\bigskip

\centerline{\bf ABSTRACT}
\begin{quotation}
Interpretation of exact results on the low-energy
limit of $4d$ $N=2$ SUSY YM  in the language of $1d$ integrability
theory is reviewed.
The case of elliptic Calogero system, associated with the flow
between $N=4$ and $N=2$ SUSY in $4d$, is considered
in some detail.
\end{quotation} \end{titlepage} \clearpage

\newpage
\tableofcontents  \newpage

\setcounter{footnote}{0}

\section{Generalities}

One of the newly emerging paradigmas of  modern quantum theory
is identification of  {\it exact} effective actions (EA) and (generalized)
integrable systems \cite{UFN}.\footnote{
(Wilsonian) effective action is a functional of background fields and
coupling constants. From the point of view of integrable systems
(defined in terms of representation cathegories of Lie groups)
these are identified as ''moduli'' and ''time''-variables respectively,
while EA is appropriate $\tau$-function. It is classical integrability that
plays role here.  ''Quantum integrability''  in such context is considered
as an example of the ''classical'' one, associated with quantum groups.
It is actually be important for description of generic (not just low-energy)
effective actions, when background fields (moduli) - and thus EA themselves  -
are operator-valued and subjected for further averaging over ''slow''
quantum fluctuations.
}
The {\it motivation} for this comes from the large reparametrization
symmetry peculiar to EA, and is rather general.
However, until recently too few examples were
known of exactly evaluated EA, and most of {\it evidence} to confirm this view
was emerging from the study of matrix models, which are at best
$1+1$-dimensional quantum  theories.
Recently, in a brilliant breakthrough, N.Seiberg and E.Witten \cite{SW,SW2}
evaluated   exact low-energy effective actions in certain
($N=2$ SUSY YM + matter) $3+1$ dimensional theories.
With no surprise, it was demonstrated in \cite{Go} that the {\it answers}
from \cite{SW,SW2} possess natural
interpretation in terms of integrable systems.

What is impressive, in this particular context just
the old-known types of integrable systems, associated with finite-\-dimensional
Lie algebras (rather than with affine and multi-loop ones)
appeared relevant. It was suggested in \cite{Go} that the reason for this
can be the pecularity of {\it low-energy} effective actions, which are obtained
from generic EA by specific Bogolubov-\-Whitham averaging procedure,
invloving a limit
$\frac{{\rm normalization\ point}}{{\rm all\ non-vanishing\ masses}}
\ \longrightarrow \ 0$.
As usual, after a limit is taken, different original models can get into the
same universality class and, presumably, a large variety of  such
classes is represented by Whitham theories on spectral surfaces of
complex dimension {\it one}.

It deserves mentioning that so far the theory of effective actions, as
well as of their low-energy limits, was never
addressed in a general context,- apart from the study of particular
models. Therefore one should hardly be surprised that the usual
intuition does not help here, and some strangely-looking phenomena
occur (like emergency of Riemann surfaces in the study of dynamics of
$4d$  theories). This subject deserves more attention and hopefully
{\it dynamical mechanisms} responsible for the Seiberg-Witten
prescriptions will once be brought to light in these investigations.

The purpose of this paper is to review the existing evidence for
identification of the Seiberg-Witten description with the data from
integrability theory on the lines of \cite{Go} and numerous subsequent
papers \cite{MW1,NT,ET,DW,M,GoM,MW2}.

\section{$4d$ Physics }

According to \cite{SW,SW2} the low-energy description of any model
of $4d$ $N=2$ SUSY {\it YM + matter} theory with the gauge group $G$
is encoded in the following data:

\bigskip

1) Riemann surface (complex curve) ${\cal C}_G(h_k | \tau)$;

2) Integrals $S_C = \oint_C dS $ along non-contractable
contours (1-cycles) on this surface.

\bigskip

This data is enough to reconstruct exact low-energy effective action.
The whole pattern depends on the moduli, $h_k$, $\ k = 1\ldots r_G$
(if there are no matter supermultiplets, otherwise their masses should
also be included), which are interpreted as vacuum
expectation values (v.e.v.) $h_k = \frac{1}{k}\langle {\rm Tr} \Phi^k\rangle$
of the scalar component $\Phi$ of $N=2$ gauge supermultiplet.
Parameter $\tau$ is the ultraviolet (UV) {\it bare} complex coupling
constant  $\tau = \frac{4\pi i}{e^2} + \frac{\theta}{2\pi}$, which has
direct physical meaning if the theory is UV finite.  As soon as the
gauge group is simple, there is only one gauge coupling constant
$\tau$ in the UV.

The simplest example of an UV finite $4d$ theory
- on which we concentrate in what follows - is the $N=2$ gauge theory
with a matter hypermultiplet with mass $m$ in adjoint of the gauge group,
which in the UV possesses $N=4$ supersymmetry and becomes
conformally invariant. In the infrared (IR) limit the gauge group
$G$ is generically broken by the v.e.v.'s $\langle \Phi \rangle$
down to $U(1)^{r_G}$. What survives are the $r_G$ light abelian
gauge supermultiplets belonging to  Cartan  part
of original supermultiplet in the adjoint of $G$.
Their scalar components parametrize the flat valleys, and the
corresponding  background fields are refered to as $a_k$.

Exact effective action of this low-energy  abelian theory is
fully described by the  ''prepotential'' ($N=2$ superpotential)
${\cal F}(a_k)$, which of course depends non-trivially also on
$h_k$ and $\tau$. One usualy introduces in addition to $a_k$
the ''dual'' variables
$a^D_k \equiv \frac{\partial{\cal F}}{\partial a_k}$ and effective
abelian charges ${\cal T}_{ij} = \frac{\partial a^D_i}{\partial a_j} =
\frac{\partial^2{\cal F}}{\partial a_i\partial a_j}$.
Moreover, whenever $m\neq 0$, at  low energies a dynamical
transmutation takes place and $\tau$ is substituted by the massive
parameter\footnote{
$b_1$ stands for the first (and unique in the case of a SUSY
YM under consideration) coefficient of (Wilsonian) $\beta$-function,
which in the $N=2$ theory is known to be equal to twice the dual Coxeter
number of $G$: $b_1 = 2{\rm h}_G^\vee$.
One can make use of conformal invariance to set $m=1$,
then the flow from $N=4$ to $N=2$ (i.e. from $m=0$ to $m=\infty$)
is described as that to  $\tau \rightarrow +i\infty$.
According to \cite{SW2} the moduli $h_k$ can be different in the
UV (in the $N=4$ theory above $m$) and in the IR (in the $N=2$ one
below $m$). This is obvious, once the moduli space in the IR
has singularities at $\Lambda_{{\rm QCD}}$-dependent points,
with $\Lambda_{{\rm QCD}}$ which itself depends on $\tau$
according to (\ref{mtscaling}). The $\tau$-dependent difference
between $h_k^{N=4}$ and $h_k^{N=2}$ is nicely described in
the framework of integrability theory, see s.\ref{CtoTC}  below.
}
 \be
\Lambda_{{\rm QCD}} ^{b_1}=
\lim_{\stackrel{\tau \rightarrow i\infty}{m \rightarrow \infty}}
m^{b_1} e^{2\pi i\tau}
\label{mtscaling}
\ee
If the set of matter multiplets in the
$N=2$ SUSY theory is not adjusted to make it UV finite,
then $\Lambda_{{\rm QCD}}$ enters description of
${\cal C}(h_k | \Lambda_{{\rm QCD}})$ instead of $\tau$.

Since both background fields $a_k$ and the v.e.v.'s $h_k$ are
some coordinates along the valleys, they are related. Relation
is non-trivial, because matrix elements (of which v.e.v. is an
example) depends non-trivially on background fields.
The main claim of \cite{SW,SW2} is that this relation can be
described by the formulas
$$
a_i = S_{A_i} = \oint_{A_i} dS, \ \ \ \
a^D_i =  S_{B_i} = \oint_{B_i} dS, \nn
$$
where $A_i$ and $B_i$ are conjugate $A$ and $B$-cycles
on the curve ${\cal C}$.\footnote{In the ''$4d$ physical'' context  the
possibility to represent explicitly (the cohomology class of) $dS$
depends largely on the adequate parametrization of the family
${\cal C}(h_k)$. For  many of the $N=2$ theories this was achieved
in refs.\cite{KLTY,AF,HO,DS,BL}  and nicely summarized in \cite{MW1}.
Sometime the number of  $A$ and $B$ cycles on ${\cal C}$ exceeds
$2r_G$, however with the proper $dS$ all the extra integrals vanish
due to the symmetries of the problem.
Of course the adequate description arises automatically in
the context of integrability theory: see \cite{Go,MW1,DW,GoM}
and section \ref{examples} below.
}
What  is not quite usual, in these theories the background fields
$a_i$, $a^D_i$ are directly measurable. According to  \cite{OW}  they enter
expression for the central charge of the $N=2$ superalgebra (central
charges, as all the other Schwinger terms, are normally operator-valued
beyond two dimensions),  $Z = \sum_i (n_ia_i + n^D_ia^D_i)$,
which defines the masses of BPS-saturated states in  the
{\it small} representations of superalgebra: ${\cal M}_C = 2|Z|_C =
2|\oint_C dS|$, with $C = \sum_i (n_iA_i + n^D_iB_i)$.

\section{Integrability theory}

According to \cite{Go}-\cite{MW2} the picture, as described
in the previous section, has exact counterpart in the theory of
integrable systems \cite{OP,Kr,Hi}. On this side, instead of fixing
a model of $N=2$ SUSY {\it YM + matter}, one inputs some integrable
system. It is represented by the Lax operator $L(z|\tau)$, which
depends on the complex ''spectral parameter'' $z$ - a coordinate on a
complex ''{\it bare} spectral curve'' , which is either a punctured Riemann
sphere $S^2$ or torus (elliptic curve) $E(\tau)$.
Then:

\bigskip

1) The entire spectral curve ${\cal C}(h_k|\tau)$ is described\footnote{
Let us remind that the main idea of algebro-geometrical approach to
integrable equations \cite{Kr} is to interpret the obvious ''time''-invariance
of eq.(\ref{CfromL}) (which follows immediately from the Lax equation,
$\frac{\partial L}{\partial t_{\alpha}} = [ L, A_{\alpha}]$,
 $\ \ A_{\alpha} =
{\cal R}(\Delta^{\alpha -1} A)$) as invariance of the (moduli of the)
spectral curve ${\cal C}$: moduli are integrals of motion. Clearly this
statement (formulated as invariance of entire exact spectral surface of
a theory under Hamiltonian flows) is not restricted to the field of
conventional integrable systems (of KP/Toda family).
Note that the curve, as defined in (\ref{CfromL}) depends also on
representation $R$ of $G$ (since $L$ belongs to some $R$, not obligatory
irreducible);  in what follows we assume that  $G$  labels the pair
(Lie group, its representation).
}
as a ramified covering over $E(\tau)$:
\be
\det \left(L(z) - t\right) = 0.
\label{CfromL}
\ee

2)  The $S_C$-variables are given by
\be
S_C = \oint_C t\omega.
\label{Sfromtom}
\ee

\bigskip

In this picture $h_k$ (which were averages $\frac{1}{k}\langle {\rm Tr}\Phi^k
\rangle$ in $4d$) are the values of Hamiltonians  (integrals of motion) of the
integrable system.
%\be
%h_k = H_k.
%\label{h=H}
%\ee
$S_C$ are its action integrals, $\frac{\partial S_C}{\partial h_k}$
being the (complex) periods of motion along the closed trajectories.
Finally, $\omega(z)$ is canonical
holomorphic 1-differential on $E(\tau)$ (''canonical'' means that
$\oint_{{\cal A}}\omega = 1$, then $\oint_{{\cal B}} \omega = \tau$).  As $\tau
\rightarrow i\infty$, $E(\tau)$ degenerates into Riemann sphere
$S^2_{(2)}$ with two punctures (conventionally placed at $0$ and $\infty$),
while $\omega$ acquires first-order poles at punctures with the
residues $\oint_{\pm{\cal A}}\omega = \pm 1$, i.e. $\omega \rightarrow
\frac{1}{2\pi i}\frac{dz}{z}$, and eq.(\ref{Sfromtom}) turns into
\be
S_C \stackrel{\tau \rightarrow i\infty}{\longrightarrow}
\frac{1}{2\pi i}\oint_C td\log z.
\label{Sfromtom'}
\ee
It is a certain $\tau$-dependent combination of the Hamiltonians
$h_k$ ($h_k^{N=4}$),
that has a smooth limit  (to be identified as $h_k^{N=2}$) as
$\tau \rightarrow i\infty$.

According to \cite{Go} the ''time''-flows of integrable theory
should be interpreted as renormalization flows (dependence on the
bare UV couplings)
in original ''target-space'' ($4d$) model.
The ''time''-\-{\it in}dependence of the
Hamiltonians, $\frac{\partial h_k}{\partial t_\alpha} = 0$,
is then nothing but
renorminvariance of physicaly sensible v.e.v.  $h_k$. Their ($h_k$)
dependence  on the low-energy effective coupling constants
(which is of course non-trivial) is then described in terms of
effective low-energy
Witham dynamics \cite{Wh},
deducible from dynamics of original integrable system. In certain sense the
Whitham procedure can be considered as quantization of original (integrable)
system around non-trivial background solutions (sometime the whole
approach is called non-linear WKB).
Such view is usefull for establishing connections to
quantum groups and Langlands duality.

After the Whitham ''slow'' time-variables $T_\alpha$ are introduced,
the prepotential ${\cal F}(a_i, T_\alpha)$ is identified as a
''quasiclassical $\tau$-function'', familiar to many from the studies
of topological field theories \cite{D,Kr2,LP} and matrix models
\cite{KMMM/MPLA}.
${\cal F}$ is quadratic function of $a_i$ and $T_\alpha$, with coefficients
made from the moduli $h_k$. However, $h_k$ are themselves
depending on $a_i$ and $T_\alpha$, what makes ${\cal F}$ a highly
non-trivial function. Still, as any quasiclassical
$\tau$-function, it is homogeneous of degree 2 (though not just
quadratic),
$$
\left(\sum_{i=1}^{r_G} a_i\frac{\partial}{\partial a_i} +
\sum_{\alpha = 0}^\infty T_\alpha\frac{\partial}{\partial T_\alpha}\right)
{\cal F} = 2{\cal F}.
$$
Within the logic of \cite{Go} this condition is nothing  but
immediate corollary of scale invariance (a part of renorminvariance)
of the v.e.v.  $h_k$,
$$
\left(\sum_{i=1}^{r_G} a_i\frac{\partial}{\partial a_i} +
\sum_{\alpha = 0}^\infty T_\alpha\frac{\partial}{\partial T_\alpha}\right)
h_k = 0.
$$
Of course, ${\cal F}$ is not homogeneous as a function of $a_i$'s alone,
because the coupling constants, not only background fields, are
scale-dependent (whenever some $\beta$-functions are non-vanishing) -
see \cite{NT},\cite{ET}  and \cite{STY}  for detailed discussion of this point
(for relation to a more sophisticated subject of Picard equations on
the moduli space of spectral curves see \cite{Pic}).  The theory of
prepotential and quasiclassical $\tau$-functions is a separate big
issue (relevant far beyond $N=2$ SUSY YM) and will be discussed
elsewhere. See \cite{NT} for a nice introduction and some references.

\section{Examples \label{examples}}

In the remaining part of these notes we demonstrate a little more explicitly
how eqs.(\ref{CfromL}) and (\ref{Sfromtom}) for Calogero family of integrable
systems can be used to reproduce the description
of refs.\cite{SW2} and \cite{DW} of the $N=4 \longrightarrow N=2$
flow. This can be helpfull to complement the
presentation of refs.\cite{Go,MW1,DW,M,GoM}. \footnote{
Some of  the explicit formulas below are given only for the fundamental
representation of $GL(N)$, generalization to other cases is always
straightforward. According to \cite{MW1}, the adequate description
of $4d$ gauge theory with non-simply laced gauge group $G$ is
in terms of integrable system, associated with the dual group $G^\vee$.
}

\subsection{Free particles (classical module space)}

Let us begin from the case when integrable system is just
that of free particles. This example will be
also used to fix the notation.
Degrees of freedom are coordinates ${\vec q} = \{q_i\}$ and
momenta ${\vec p} = \{p_i\}$. The bare spectral surface (where the
spectral parameter $z$ originally takes values) is Riemann
sphere $S^2_{(2)}$. In order to write down the Lax operator, let
us denote the generators of $G$, associated with Cartan
subalgebra and with the roots $\pm {\vec \alpha}$ through
${\vec H}$ and $E_{\pm{\vec \alpha}}$ respectively. The ''affine
root'' will be denoted ${\vec \alpha}_0$.

The Lax operator (familiar for many from the Drinfeld-Sokolov
construction) is
$$
L_{\rm free}(z) = {\vec p}{\vec H} + \sum_{{\rm simple}\
{\vec \alpha} > {\vec 0}} E_{{\vec \alpha}}  +  zE_{{\vec \alpha}_0}.
$$
In the fundamental representation of $GL(n)$
$$
L_{\rm free}^f(z) = \left(\begin{array}{cccccc}
p_1  & 1      &      0  &       &  0  &  0 \\
0      & p_2  &      1  &       &  0  &  0  \\
0       &   0     &   p_3  & \ldots  &  0 & 0 \\
 & & \ldots & & &  \\
0 & 0 & 0 &  & p_{n-1} & 1 \\
z & 0 & 0 & & 0 & p_n
\end{array} \right)
$$
The Hamiltonians (integrals of motion) are
\be
h_k = h^{(0)}_k \equiv \frac{1}{k} \sum_{i=1}^n p_i^k.
\label{h0k}
\ee

The full spectral curve ${\cal C}_{\rm free}(h_k)$ is given by (\ref{CfromL}):
$$
{\cal C}_{{\rm free}}(h_k): \ \ \ \ \ \
\det \left(t - L_{\rm free}(z)\right) = 0,
$$
or, for the fundamental representation of $GL(n)$,
$$
\prod_{i=1}^n (t - p_i)  = z.
$$
In other words, $p_i$ are  coordinates on the
''classical module space'' in the terminology of \cite{SW,SW2}.

Now we spend a paragraph for a play with notation.
Let us denote
$$
\hat P_G(t |h) \equiv
\left.\det\left(1 - t^{-1}L_{\rm free}\right)\right|_{z=0}.
$$
Since in $4d$ picture $h_k = \frac{1}{k}\langle {\rm Tr}\Phi^k\rangle$,
this can be also defined as
$$
\hat P_G(t|h) = \ \langle \det(1 - t^{-1}\Phi) \rangle.
$$
$\hat P_G$ is obviously a polynomial in $t^{-1}$,
$$
\hat P_G(t|h) = \sum_k s_k^G(h) t^{-k},
$$
its coefficients being Schur polinomials of $h$'s.
In the case of fundamental representation of $GL(n)$ -
which we use in all the illustrative examples -
it  is more convenient to deal with
$P_n(t) = t^n\hat P(t) = \left.\det\left(t -
L_{\rm free}\right)\right|_{z=0} =
\ \langle \det (t - \Phi) \rangle $. Thus
\be
P_n(t|h) = \sum_{k=0}^n s_k^f(h) t^{n-k},
\label{defPn}
\ee
where the fundamental Schur polinomials are defined by
$e^{\left(-\sum_{k=0}^\infty h_kt^{-k}\right)} =
\sum_{k=0} s_k^f(h) t^{-k}$. (Note, that while $h_k \neq 0$
even for $k >n$ they depend algebraically on the first $n$
ones, $h_1,\ldots,h_n$. For {\it such} set of $\{h_k\}$,
all the Schur polinomials $s_{k>n}^f(h) = 0$. Their other
characteristic property is
$\frac{\partial s_k^f(h)}{\partial h_l} = s_{k-l}^f(h)$.)

In the new notation, the curve ${\cal C}_{\rm free}(h)$
for the case of $GL(n)$ becomes
\be
{\cal C}_{{\rm free}}(h_k): \ \ \ \ \ \
P_n(t|h) = z,
\label{freeC}
\ee
with $P_n$ given by (\ref{defPn}).
For generic integrable systems, there is no such
spliting between $t$ and $z$ variables: instead
one can express eq.(\ref{CfromL}) through a more sophisticated
$t$-polinomial
\be
{\cal P}_n(t|z|h) = \sum_{k=0}^n s_k^f(h) {\cal T}_{n-k}(t|z),
\ee
where ${\cal T}_k(t|z)$ are still $k$-th order polinomials in $t$,
but not just $t^k - z\delta_{kn}$ as in (\ref{freeC}): they
depend non-trivially on the spectral parameter $z$.\footnote{
Examples of this phenomenon below will include
rational and elliptic Calogero systems, for which the bare spectral
curve is $S^2_{(1)}$ and $E(\tau)$ respectively. Naturally, in the latter
case the polinomials of Donagi and Witten \cite{DW} are reproduced
(or rather linear combinations of those, which we believe are more
appropriate, see s.\ref{DWcomp} below).
}

To complete our discussion of the free particle model, let us mention
that since $\frac{1}{2\pi i}td\log z =
\sum_{i=1}^n \frac{t}{t-p_i}\frac{dt}{2\pi i}$,
the integrals  $S_C$ are just linear combinations of $p_i$. Since one does
not expect any quantum corrections, this is a right answer: $a_i $
just coincide with the moduli $p_i$. This is also in agreement with the
quasiclassical interpretation of (\ref{Sfromtom}): for a free particle
the normalized action integral along a closed trajectory of the length $l$
is $S = \frac{1}{l}\oint pdq =
\frac{1}{l}\oint p\dot q dt = \frac{p^2T}{ml}$, while the period
is $T = \frac{l}{\dot q} = \frac{lm}{p}$, so that $S = p$.

\subsection{Toda-chain system (quantum module space in the
absence of matter) \label{TCsec}}

The bare spectral curve for this model is still $S^2_{(2)}$,
and the Toda-chain Lax operator, associated with the algebra
$G$ is given by \cite{OP}:
\be
L_{\rm TC} = {\vec p}{\vec H} +
\sum_{{\rm simple}\  {\vec \alpha} > {\vec 0}}
\left(E_{{\vec \alpha}}  +
e^{{\vec \alpha}{\vec q}}E_{-{\vec \alpha}}\right) +
zE_{{\vec \alpha}_0} + \frac{1}{z} e^{{\vec \alpha}_0{\vec q}}
E_{-{\vec \alpha}_0}.
\label{TCLax}
\ee

In the fundamental representation of $GL(n)$ the roots are represented
as matrices $E_{ij}$ with  non-vanishing entries at the crossing
of $i$-th row and $j$-th column.  For positive roots $i <j$ (upper tringular
matrices), for negative roots $i>j$. Diagonal matrices represent Cartan
elements. The simple positive/negative roots belong to the first
upper/lower subdiagonal, the affine roots $\pm {\vec \alpha}_0$ are
located at the left lower/ right upper corner respectively. Thus
$$
L_{\rm TC}^f(z) = \left(\begin{array}{cccccc}
p_1  & 1      &      0  &       &  0  &  \frac{1}{z}e^{q_1-q_n} \\
e^{q_2-q_1}      & p_2  &      1  &       &  0  &  0  \\
0       &   e^{q_3 - q_2}  &   p_3  & \ldots  &  0 & 0 \\
 & & \ldots & & &  \\
0 & 0 & 0 &  & p_{n-1} & 1 \\
z & 0 & 0 & & e^{q_n-q_{n-1}}& p_n
\end{array} \right)
$$
As soon as $z^{-1}\neq 0$, the system is {\it periodic} Toda chain,
i.e. one can write $q_{n+1} \equiv q_1$.
The first Hamiltonians (integrals of motion) are
\be
h_1 = \sum_{i=1}^n p_i, \nn \\
h_2 = \sum_{i=1}^n \left(\frac{1}{2}p_i^2 + e^{q_{i+1} - q_i}\right) =
h_2^{(0)} + \sum_{i=1}^n e^{q_{i+1} - q_i}, \nn \\
\ldots
%h_k = H_k = \frac{1}{k} \sum_{i=1}^n p_i^k.
\ee
If one consdiers $SL(n)$ group instead of $GL(n)$, the
first Hamiltonian $h_1 = 0$. Looking at the second Hamiltonian
one easily recognizes the simplest equation of motion for the Toda chain:
$$
\frac{\partial^2 q_i}{\partial t_2^2} = e^{q_{i+1}-q_i} - e^{q_i-q_{i-1}}.
$$
It is now an easy calculation to find the curve
${\cal C}_{\rm TC}(h_k)$ from eq.(\ref{CfromL}):
\be
0 = \det \left(t - L_{\rm TC}^f(z)\right) = \nn \\
= t^n - t^{n-1}\left(\sum_{i=1}^n p_i\right)  +
t^{n-2}\left( \sum_{i<j}^n p_ip_j - \sum_{i=1}^n e^{q_{i+1}-q_i}\right)  +
\ldots - z - \frac{1}{z}  = \nn \\
= t^n - h_1t^{n-1} + \left(\frac{h_1^2}{2} - h_2\right) t^{n-2} + \ldots -
z - \frac{1}{z} = \nn \\
= P_n(t|h) - \left(z + \frac{1}{z}\right),
\ee
where $P_n(t|h)$ is just the polinomial  (\ref{defPn}) from the previous
section. There are only two differences from the free-particle case:
First, the coefficients of this $P_n(t)$ are now Schur polinomials of
Toda-chain Hamiltonians $h_k$,
not just of the free ones $h^{(0)}_k$  - but this has no
effect on the {\it shape} of  the curve ${\cal C}$. Second - and this
is the only trace of the switch from one integrable system to another -
is that instead of (\ref{freeC}) we now get
\be
{\cal C}_{TC}(h):\ \ \ \ \ \  P_n(t|h) = z + \frac{1}{z}
\label{TCC}
\ee
with just the {\it same} polinomial $P_n(t|h)$.

A few comments are in order before we end this section.
By a change of variables $z - \frac{1}{z} = Y$, eq.(\ref{TCC})
can be transformed to the form
$$
Y^2 - \left(P_n(t|h)\right)^2 = \frac{1}{4},
$$
which was suggested for description of the quantum module space
for the $N=2$ gauge theory without matter in
refs.\cite{KLTY,AF,HO}.
Since now $z = \frac{1}{2} (Y + P_n(t|h))$, the integrals
$$
2\pi i S_C = \oint_C td\log z = \oint_C t  d\log(Y + P_n(t|h)),
$$
in agreement with \cite{KLTY,AF,HO}.

In the simplest case of $G = SL(2)$, an obvious substitution
$z = -e^{i\varphi}$ describes ${\cal C}_{TC}$
as $t^2 - h^2 + 2\cos \varphi = 0$,
and
$$
2\pi S_C = \oint_C td\varphi = \oint_C\sqrt{h_2 - 2\cos\varphi}\ d\varphi,
$$
where the two integrals with $C=A$ and $C=B$ are over allowed and
forbidden zones of classical motion in the sine-Gordon potential.
This expression is the most straightforward illustration for the
quasiclassical  nature of (\ref{Sfromtom}), if considered
from the point of view of integrability theory (while for $4d$ physics
this is a low-energy Bogolubov-Whitham limit).

Eq.(\ref{TCC}) can be also represented as a system
\be
z_+ + z_- = P_n(t|h),  \nn \\
z_+z_- = 1
\ee
The spectral curve ${\cal C}_{\rm TC}$ is obtained by factorization
w.r.to $Z_2$ transformation $z_+ \leftrightarrow z_-$.
The singular (orbifold) points $z_+ = \pm 1$ are exactly the
two punctures of the bare spectral surface $S^2_{(2)}$.
These orbifold points are blowed up in transition to
$E(\tau)$ and ${\cal C}(h_k|\tau)$

\subsection{Calogero system}

\subsubsection{Elliptic Calogero (quantum moduli space in the UV-finite case)}

Expression (\ref{TCLax}) for the Toda-chain Lax operator looks
a little artificial from the algebraic point of view, because it
gives a special role to the simple roots. Clearly, its ''more symmetric''
counterpart exists, and is known as Calogero system \cite{OP} (its
full name would be the Sutherland-\-Calogero-\-Moser-\-Olshanetsky-\-
Perelomov-\-Ruijsenaars  system). The Lax operator is
\be
L(z) = {\vec p}{\vec H} + \sum_{{\vec \alpha}} F({\vec q}{\vec \alpha}|z)
E_{{\vec \alpha}},
\ee
where the sum goes over all (positive and negative) roots of $G$.
Generically the bare spectral curve is elliptic, $E(\tau)$, and
the function $F(q|z)$ is certain elliptic function of $z$ (it actually
is not quite a double periodic {\it function}, but belongs to a linear
bundle over $E(\tau)$).
Its explicit appearence depends on parametrization of $E(\tau)$:
in varience with the Riemann sphere, elliptic curve has at least two
widely useful  parametrizations, to be refered to as ''flat''
and ''elliptic''. In the flat parametrization $E(\tau)$ is represented as
a parallelogramm with the sides $1$ and $\tau$, and we denote the
spectral parameter in this parametrization through $\xi$ (leaving
notation $z$ for parameter on $S^2$, which arises in degenerations of
$E(\tau)$). Elliptic parametrization represents the same $E(\tau)$ as
a double covering over Riemann sphere,
$y^2 = (x - e_1(\tau))(x - e_2(\tau))(x - e_3(\tau))$.

%Now we can not avoid going into some technical details.
Peculiar for the flat parametrization is a family of Weierstrass
elliptic functions, which can be all constructed from theta-functions
and represented as infinite sums or products.
The double periodic Weierstrass function
{\it per se},
$$
\wp (\xi) = \frac{1}{\xi^2} +
\sum_{\stackrel{-\infty < N,M <\infty}{N^2+M^2\neq 0}}
\left(\frac{1}{(\xi + Nw_1 + Mw_2)^2} -
\frac{1}{(Nw_1+Mw_2)^2}\right),\ \ \  \
\tau = \frac{w_2}{w_1},
$$
is even and has double pole at $\xi = 0$. We also need its integrals,
$\zeta(\xi)$ and $\sigma(\xi)$:
$\wp (\xi) = -\frac{d\zeta(\xi)}{d\xi} =
-\frac{d^2\log\sigma(\xi)}{d\xi^2}$. They are not double periodic
(but acquire non-trivial phase factors under the shifts
$\xi \rightarrow \xi + w_l$, $l=1,2$),
both are odd, $\zeta$ has simple pole,
and $\sigma$ - simple zero at $\xi = 0$.
Further details can be found in any textbook on elliptic functions \cite{AS}.

Generic Calogero function $F(q|\xi)$ is expressed
through $\zeta$ and $\sigma$ \cite{OP}:
\be
F(q|\xi) = g\frac{\sigma(q - \xi)}{\sigma(q)\sigma(\xi)}
e^{q\zeta(\xi)},
\ee
It is double periodic  in $\xi$, but not in $q$. Instead
$$
F(q \pm w_l|\xi) = F(q|\xi) e^{\pm(w_l\zeta(\xi) - \zeta(w_l)\xi)}.
$$
Since exponential factor is independent of $q$, this transformation rule
immediately implies that any  product of the form
$$
F_{i_1i_2}F_{i_2i_3}  \ldots F_{i_ki_1},
$$
with $F_{ij} \equiv F(q_{ij}|\xi)$,
$q_{ij} = q_i - q_j$ is double periodic in all the $q_i$'s
and $\xi$. In particular,
\be
F(q|\xi)F(-q|\xi) = g^2(\wp(\xi) - \wp(q)).
\label{2-prod}
\ee
Further simplifications arise after symmetrization
over indices $i_1,\ldots,i_r$.
Introduce for $i_1<i_2 <\ldots < i_r$
\be
({\cal S}^rF)_{i_1\ldots i_r} \equiv \frac{1}{r}
\sum_{\stackrel{{\rm all\ permutations\ of }}{i_1\ldots i_r}}
F_{i_1i_2}F_{i_2i_3}  \ldots F_{i_ki_1}.
\label{Sproddef}
\ee
Every item in the sum has two simple poles in every argument $q_{i_s}$
(so that $({\cal S}^rF)$ could have simple poles
whenever $q_{i_{s}} = q_{i_{s'}}$).
It  is, however, easy to check that the residue at
the pole cancels completely
between different items in the sum (\ref{Sproddef}) as $r > 2$.
Thus, as a function
of any $q_{i_s}$ our $({\cal S}^rF)$
has no singularities - therefore it does not
depend on any of $q_{i_s}$'s at all.
This means that the indices $i_1\ldots i_r$
can be also omited at the l.h.s. of (\ref{Sproddef}), and
$({\cal S}^rF)_{i_1\ldots i_r} = ({\cal S}^rF)(\xi)$
is a function of $\xi$ only.
This function can be easily evaluated (for example, by evaluating the sum
on the r.h.s. of (\ref{Sproddef}) at some special
values of $q_{i_s}$, like $q_{k} = k\xi$):
\be
({\cal S}^rF)(\xi) = (-g)^r\left(\frac{d}{d\xi}\right)^{r-2}
\wp(\xi),\ \ \  \ r>2.
\label{r-Sprod}
\ee
(Asymptotical behaviour,  $({\cal S}^rF)(\xi) =
\frac{(r-1)!}{\xi^r}(1 + o(\xi))$
is very simple to reproduce in the rational limit of Calogero system,
see s.\ref{racal} below.)

We now return to Calogero system and restrict
to our usual example of fundamental
representation of $GL(n)$.
The first Hamiltonians of Calogero system are:
\be
h_1 = \sum_{i=1}^n p_i, \nn \\
h_2 = \frac{1}{2}\sum_ {i=1}^n p_i^2  -g^2  \sum_{i<j} \wp(q_i - q_j|\xi) =
  h_2^{(0)} - g^2  \sum_{i<j} \wp(q_{i j}|\xi), \nn \\
h_3 = h_3^{(0)} - g^2  \sum_{i\neq j} p_i\wp(q_{ij}|\xi),
\ldots
\label{calham}
\ee
(see (\ref{h0k}) for the definition of $h^{(0)}_k$).
The Lax operator
$$
L(\xi) = \left(\begin{array}{ccccc}
     p_1 & F_{12} & F_{13} & & F_{1n} \\
     F_{21} & p_2 & F_{23} &\ldots & F_{2n} \\
  &&\ldots && \\
     F_{n1} & F_{n2} & F_{n3} && p_n \end{array}\right)
$$
(Note that there is no special role assigned to the affine roots: they
appear automatically in appropriate degeneration limit
$E(\tau) \rightarrow S^2_{(2)}$, see s.\ref{CtoTC} below.)
Eq.(\ref{Sfromtom}) is now:
\be
{\cal C}(h_k|\tau): \ \ \ \ \ \ {\cal P}_n(t|h|\xi) = 0,
\label{CalC}
\ee
with
\be
{\cal P}_n(t|h|\xi) \equiv  \det \left(t - L^f(\xi)\right) = \nn \\
= t^n\ -\ t^{n-1} \sum_{i=1}^n p_i \ + \
t^{n-2}\sum_{i<j}^n \left(p_ip_j - F_{ij}F_{ji}\right) \ - \  \nn \\ - \
t^{n-3}\sum_{i<j<k}^n
\left( p_ip_jp_k - p_iF_{jk}F_{kj} - p_jF_{ik}F_{ki} -
p_kF_{ij}F_{ji} + F_{ij}F_{jk}F_{ki} + F_{ik}F_{kj}F_{ji}
\right) \ + \nn \\
+\ \ \ \ldots
\label{Pnfirst}
\ee
 The first  few terms on the r.h.s. are easily evaluated with the help of
(\ref{2-prod}) and explicit expressions (\ref{calham})
for the first few $h_k$.
The first two are just $t^n - h_1t^{n-1}$ (of course $h_1^{(0)} = h_1$).
Further:

The $t^{n-2}$ term:
$$
\sum_{i<j} p_ip_j = \frac{1}{2}\left(\left(\sum_{i=1}^n p_i\right)^2 -
\sum_{i=1}^n p_i^2\right)
= \frac{h_1^2}{2} - h_2^{(0)};
$$
$$
- \sum_{i<j} F_{ij}F_{ji} \ \stackrel{(\ref{2-prod})}{=}\
-\frac{n(n-1)}{2}g^2\wp(\xi) +
g^2\sum_{i<j} \wp(q_{ij}).
$$
Together, the last items of the two expressions combine into
the full Hamiltonian
$-h_2$, and we get:
$$
\left(\frac{h_1^2}{2} - h_2 - \frac{n(n-1)}{2}g^2\wp(\xi)\right) t^{n-2}
$$

The $t^{n-3}$ term:
$$
\sum_{i<j<k} p_ip_jp_k = \frac{h_1^3}{6} - h_1h_2^{(0)} + h_3^{(0)};
$$
\be
- \sum_{i<j<k} (p_iF_{jk}F_{kj} +  p_jF_{ik}F_{ki} + p_kF_{ij}F_{ji} )
= \nn \\
\ \stackrel{(\ref{2-prod})}{=}\
-\frac{n(n-1)}{2}g^2\wp(\xi)\left(\sum_{i=1}^n p_i \right)\ +\
g^2\sum_{i<j<k} \left(p_i\wp(q_{jk}) + p_j\wp(q_{ik} )+
p_k\wp(q_{ij})^{\phantom .}
\right ) = \nn \\
= -\frac{n(n-1)}{2}g^2\wp(\xi) h_1\ +\
g^2\left(\sum_{i=1}^n p_i\right)\left(\sum_{j<k} \wp(q_{jk})\right) -
g^2\sum_{i\neq j}p_i\wp(q_{ij}).
\nn
\ee
The two last terms serve to complement $h_2^{(0)}$ and $h_3^{(0)}$ till full
$h_2$ and $h_3$. The remaining contribution to the $t^{n-3}$ term is just
$$
\sum_{i<j<k} ({\cal S}^3F)_{ijk}  \
\stackrel{(\ref{r-Sprod})}{=}\ \frac{n(n-1)(n-2)}{6}
({\cal S}^3F)(\xi) = -\frac{n(n-1)(n-2)}{6}g^3 \wp'(\xi).
$$

Bringing things together we obtain for (\ref{Pnfirst}):
\be
{\cal P}(t|h|\xi) = t^n\ - \ h_1t^{n-1} \ + \
\left(\frac{h_1^2}{2} - h_2 -
\frac{n(n-1)}{2}g^2\wp(\xi)\right) t^{n-2} \ + \nn \\
+\ \left( \frac{h_1^3}{6} - h_1h_2 + h_3 - \frac{n(n-1)}{2}g^2\wp(\xi)h_1 -
\frac{n(n-1)(n-2)}{6}g^3 \wp'(\xi)\right) t^{n-3} \ - \ \ldots
\ee
It is reasonable to rearrange this sum, collecting terms with the same
$h$-dependence (and thus in a sense separate $h$- and $\tau$ dependencies -
the latter one  entering through  Weierstrass functions):
\be
{\cal P}_n(t|h|\xi) = \sum_{k=0}s_k^f(h) {\cal T}_{n-k}(t|\xi) =
\label{PthroughT} \\
= {\cal T}_n(t|\xi) - h_1{\cal T}_{n-1}(t|\xi) +
\left(\frac{h_1^2}{2}-h_2\right){\cal T}_{n-2}(t|\xi) -
\left(\frac{h_1^3}{6} - h_1h_2 + h_3\right){\cal T}_{n-3}(t|\xi) + \ldots
\ee
The new $h$-{\it independent} $t$-polinomials here are:
$$
{\cal T}_n(t|\xi) = t^n - \frac{n(n-1)}{2}g^2\wp(\xi)t^{n-2} +
\frac{n(n-1)(n-2)}{6}g^3\wp'(\xi) t^{n-3} - \ldots
$$
Our next task is to derive a generic formula for ${\cal T}_n$.

The coefficient of $t^{n-\ell}$ in ${\cal T}_n$ comes from evaluation of
the minor of the Lax operator at $p_i = 0$,
$$
M_\ell =  \left(\begin{array}{ccccc}
      0 & F_{12} & F_{13} & & F_{1\ell} \\
     F_{21} & 0 & F_{23} &\ldots & F_{2\ell} \\
  &&\ldots && \\
     F_{\ell1} & F_{\ell 2} & F_{\ell 3} && 0 \end{array}\right)
$$
$M_\ell$ is obviously an algebraic combination of $({\cal S}^rF)$:
$$
M_2 = -({\cal S}^2F)_{12}, \ \ \
M_3 = +({\cal S}^3F)_{123},
$$
$$
M_4 = ({\cal S}^2F)_{12}({\cal S}^2F)_{34} +
({\cal S}^2F)_{13}({\cal S}^2F)_{24} +
    ({\cal S}^2F)_{14}({\cal S}^2F)_{23} \ - \
({\cal S}^4F)_{1234}, \ \ \ \ \ldots
$$
The $q$-dependent part of $M_\ell$ is always absorbed into the Hamiltonians
$h_k$ and does not contribute to the coefficients
of $h$-independent ${\cal T}_n$.
Thus in the study of ${\cal T}_n$ we
can neglect $q$-dependent contributions
to $M_\ell$ and substitute $M_{\ell}(\{q\};\xi)
\rightarrow \hat M_{\ell}(\xi)$.
According to (\ref{r-Sprod}) the only source of
$q$-dependence is $({\cal S}^2F)$,
thus for our purposes it is enough to drop the second item at the r.h.s.
of (\ref{2-prod}),
i.e. just use the same expression (\ref{r-Sprod}) for $r=2$.
After this the matrix indices can be neglected, and we face a pure
combinatorial problem of  decomposing an integer
$\ell$ into a sum of different
integers $r_s$, every $r_s$ coming with multiplicity
$m_s$. Thus
$$
\hat M_\ell =  (-)^\ell
\sum_{\stackrel{\stackrel{2\leq
r_1 < r_2 < \ldots}{m_s > 0}}{\sum_s m_sr_s = \ell}}
\frac{\ell !}{\prod_s m_s!(r_s!)^{m_s}} \prod_s (-{\cal S}^{r_s}F)^{m_s}.
$$
It remains to include the factor
$\frac{n!}{\ell !(n- \ell)!}$ for the number of minors
like $M_\ell$ in the $n\times n$ matrix,
and substitute $({\cal S}^rF)$ from (\ref{r-Sprod}) in order to obtain:
\be
t^{-n}{\cal T}_n(t|\xi) = 1 \ + \  \nn \\ +
\sum_{\stackrel{2\leq r_1 < r_2 < \ldots}{m_s > 0}}
\frac{n!}{ \left(n - \sum_s m_sr_s\right)!}
\prod_s \frac{(-)^{m_s}}{m_s!(r_s!)^{m_s}}
\left(-\frac{g}{t}\right)^{\sum_s m_sr_s}
\prod_s \left(\partial_\xi^{r_s-2}\wp(\xi)\right)^{m_s}
\label{explT}
\ee
Obviously, $\frac{\partial{\cal T}_n}{\partial t} = n{\cal T}_{n-1}$ -
what makes ${\cal T}_n(t)$ somewhat similar to Schur polinomials.

Equations  (\ref{CalC}), (\ref{PthroughT}) and (\ref{explT})
provide a complete
description of the full spectral curve ${\cal C}(h_k|\tau)$,
induced by the flat
parametrization of the bare spectral curve $E(\tau)$.
It should be supplemented
by eq.(\ref{Sfromtom}) for the action integrals.
Since in the flat parametrization
$\omega = \frac{1}{w_1} d\xi$,
$$
S_C = \frac{1}{w_1}\oint_C td\xi.
$$

One can easily convert to elliptic parametrization of $E(\tau)$,
when instead of
$\xi$ one uses $x,y$:  $y^2 = \prod_{a=1}^3 \left(x-e_a(\tau)\right)$.
It is enough to recall that
$$
(\wp')^2(\xi) = 4\wp^3(\xi)  - {\rm g}_2(\tau)\wp(\xi) - {\rm g}_3(\tau) =
4\prod_{a=1}^3\left(\wp(\xi) - e^0_a(\tau)\right)
$$
where $e_1^0 + e_2^0 + e_3^0 = 0$. Given $e_a(\tau)$ in generic position,
$e_a^0 = e_a - s$, $\ \ s(\tau) \equiv \frac{1}{3}\sum_a e_a$.
Thus we have identification:
\be
x = \wp(\xi) + s(\tau), \nn \\
y = \frac{1}{2}\wp'(\xi).
\label{flat/ell}
\ee
Higher derivatives of $\wp$-function are algebraic functions of
$\wp$, $\wp'$, and thus of $x, y$ and $e_a$, for example:
\be
\wp''(\xi) = 6\wp^2(\xi) - \frac{g_2}{2} =
6(x-s)^2 - \sum_{a=1}^3 (e_a^0)^2 = 6x^2 - 12xs + 9s^2 -
\sum_{a=1}^3 e_a^2, \nn \\
\wp'''(\xi) = 12\wp\wp'(\xi) = 24(x-s)y, \nn \\
\ldots
\label{wpders}
\ee
Substituting these expressions into (\ref{explT})
one obtains ${\cal T}_n(t|x,y)$
and thus the equation for the spectral curve ${\cal C}(h_k|\tau)$ in elliptic
parametrization. Comparison with description of \cite{SW2} (in the only
available case of $SU(2)$) identifies the Calogero coupling $g$ with the mass
of adjoint hypermultiplet:\footnote{
For this, the curve ${\cal C}_{SU(2)}(h_2|\tau)$, defined by the pair of
equations, ${\cal T}_2 - h_2{\cal T}_0 = t^2 - g^2(x-s) - h_2 =0$, and
$y^2 = \prod_{a=1}^3 (x - e_a)$, or just
$w^2 \equiv \left(\frac{yt}{g}\right)^2 = \left(x + \frac{h_2}{g^2} - s\right)
\prod_{a=1}^3 (x-e_a)$,
should be compared with the one from \cite{SW2},
$w^2 = \prod_{a=1}^3 \left(x - 2\hat e_ah_2 - \frac{1}{4}\hat e_a^2m^2\right)$.
The argument of \cite{SW2} requires $\hat e_a(\tau)$ to have certain
modular weight and asymptotics $\hat e_1(\tau = i\infty) = \frac{2}{3}$,
thus they are actually $\hat e_a = \left(\frac{w_1}{\pi}\right)^2e_a^0 =
\left(\frac{w_1}{\pi}\right)^2(e_a - s)$.
The two curves are equivalent provided the double ratios coincide,
$$
\frac{g^{-2}h_2 -s + e_1}{g^{-2}h_2 - s  + e_2}
\cdot\frac{e_3 - e_2}{e_3 - e_1} =
\frac{(e_3 - e_2)
\left(2h_2 + (e_2^0 + e_3^0)
\frac{m^2}{4}\left(\frac{w_1}{\pi}\right)^2\right)}{(e_3 - e_1)
\left(2h_2 + (e_1^0 +
e_3^0)\frac{m^2}{4}\left(\frac{w_1}{\pi}\right)^2\right)},
$$
i.e. $h_2 - g^2(s - e_{1,2}) = h_2 +
\frac{m^2}{8}\left(\frac{w_1}{\pi}\right)^2(e_{2,1}^0 + e_3^0) =
h_2 + \frac{m^2}{8}\left(\frac{w_1}{\pi}\right)^2(s - e_{1,2})$,
or $g = \frac{m}{2\sqrt{2}}\frac{iw_1}{\pi}$.
The $h_2$ here is by definition $h_2^{N=4}$.
\label{footSW}
}
\be
g^2 = \frac{m^2}{8}\left(\frac{iw_1}{\pi}\right)^2.
\label{gverm}
\ee
Finally, since $\omega = \frac{1}{w_1}d\xi =
\frac{1}{w_1}\frac{d\wp(\xi)}{\wp'(\xi)} = \frac{2}{w_1} \frac{dx}{y}$,
$$
S_C = \frac{2}{w_1}\oint_C \frac{tdx}{y}.
$$
Note, that in elliptic parametrization $w_1$ is not independent of $e_a(\tau)$,
and can not be put equal to unity, what is always
allowed in the flat parametrization.
Instead, $w_1 = 2\oint_{\cal A} \frac{dx}{y}$.

\subsubsection{Appendix.
Comparison with Donagi-Witten polinomials \label{DWcomp}}

Since there is no general formula for all the $\partial^k\wp(\xi)$ in elliptic
parametrization, one
can just evaluate every particulat ${\cal T}_n$ explicitly.
For this we first need them in the flat parametrization.
Allowed sets $\{r_m\}$ for the first  few $n$ in (\ref{explT}) are:
\be
\begin{array}{lc}
n=0: &\emptyset, \nn \\
n=1: &\emptyset, \nn \\
n=2: &\{(2_1)\}, \nn \\
n=3:  &\{(2_1), (3_1)\}, \nn \\
n=4:  &\{(2_1), (3_1), (4_1), (2_2)\}, \nn \\
n=5:   & \{(2_1), (3_1), (4_1), (2_2), (5_1), (2_1,3_1)\},
\end{array}  \nn \\
\ldots
\nn
\ee
Given this, we read from (\ref{explT}):
\be
\begin{array}{ll}
{\cal T}_0 = &1, \nn \\
{\cal T}_1 =  &t,  \nn \\
{\cal T}_2 = &t^2 - g^2\wp(\xi), \nn \\
{\cal T}_3 = &t^3 - 3g^2\wp(\xi) t + g^3\wp'(\xi), \nn \\
{\cal T}_4 = &t^4  - 6g^2\wp(\xi)t^2 + 4g^3\wp'(\xi)t - g^4\left(\wp''(\xi) -
3\wp^2(\xi)\right),\nn \\
{\cal T}_5 = &t^5 - 10g^2\wp(\xi)t^3 + 10g^3\wp'(\xi)t^2 -
5g^4\left(\wp''(\xi) - 3\wp^2(\xi)\right)  +
g^5\left(\wp'''(\xi) - 10\wp\wp'(\xi)\right),
\end{array}\nn \\
\ldots
\ee
Next we omit $g$ (rescale $t \rightarrow gt$ and ${\cal T}_n
\rightarrow g^n{\cal T}_n$)
and substitute (\ref{flat/ell}) and (\ref{wpders}):
\be
\begin{array}{ll}
{\cal T}_0 = &1, \nn \\
{\cal T}_1 =  &t,  \nn \\
{\cal T}_2 = & t^2 - x + s \ \nn \\ & \stackrel{s=0}{\Longrightarrow}\
t^2 - x, \nn \\
{\cal T}_3 = &t^3 - 3xt + 3st + 2y   \nn \\
    & \stackrel{s=0}{\Longrightarrow}\    t^3 - 3xt + 2y, \nn \\
{\cal T}_4 = &t^4  - 6xt^2 + 6st^2 + 8yt - 3x^2 + 6sx - 6s^2  + 3s_2
%\sum_{a=1}^3 e_a^2
\nn \\   &\ \stackrel{s=0}{\Longrightarrow}\
 t^4 - 6xt^2 + 8yt - 3x^2 + 3s_2
%\sum_{a=1}^3 e_a^2
, \nn \\
{\cal T}_5 = &t^5 -  10xt^3 + 10st^3 + 20yt^2 - 15x^2t +
30sxt - 30s^2t   + 15s_2t
%\left(\sum_{a=1}^3 e_a^2\right)
 + 4xy - 4sy \nn \\
  &\ \stackrel{s=0}{\Longrightarrow}\  t^5 - 10xt^3 + 20yt^2 -
15x^2t + 15s_2t
%\left(\sum_{a=1}^3 e_a^2\right)
+ 4xy,
\end{array}\nn \\
\ldots
\ee
Here $s = s_1(\tau) \equiv \frac{1}{3}\sum_{a=1}^3 e_a,  \ \ \ \ \
s_2(\tau) \equiv    \frac{1}{3}\sum_{a=1}^3 e_a^2$.

These expressions can  be now compared with Donagi-Witten
polinomials \cite{DW}:
\be
\begin{array}{lll}
{\cal T}_0^{\rm DW} = &1 &=\  {\cal T}_0, \nn \\
{\cal T}_1^{\rm DW} =  &t &=\  {\cal T}_1,  \nn \\
{\cal T}_2^{\rm DW}  = &t^2 - x  &=\  {\cal T}_2 - s{\cal T}_0, \nn \\
{\cal T}_3^{\rm DW} = &t^3 - 3xt + 2y  &= \ {\cal T}_3 - 3s{\cal T}_1,  \nn \\
{\cal T}_4^{\rm DW}  = &t^4  - 6xt^2 + 8yt - 3x^2 + 12sx  &= \
     {\cal T}_4 - 6s{\cal T}_2 + \left(12s^2 -  3s_2
%\sum_{a=1}^3 e_a^2
\right){\cal T}_0, \nn \\
{\cal T}_5^{\rm DW} = &t^5 -  10xt^3 + 20yt^2 -  15x^2t + & \nn \\
      & \ \ \ \ \ \ + 60sxt - 4xy - 24sy   &   =\
     {\cal T}_5 -10s{\cal T}_3 +5\left(12s^2 - 3s_2
%\sum_{a=1}^3 e_a^2
\right) {\cal T}_1,
\end{array}\nn \\
\ldots
\ee
We see that  Donagi-Witten polinomials are some linear combinations of
${\cal T}_n$, however, the linear transformation depends on $\tau$ and
thus is not quite innocent.
Moreover, this dependence does not disappear even for the natural choice
$s = \frac{1}{3}\sum_a e_a(\tau) = 0$, since $s_2(\tau) = \frac{1}{3}s^2 -
\frac{1}{6}{\rm g}_2(\tau)$ can not be made vanishing by choice.
Let us remind that ${\cal T}_n$ are defined in such a way that the
equation for the curve is
$$
{\cal C}(h_k|\tau): \ \ \ \ \   \sum_k  s_k^f(h){\cal T}_{n-k} = 0,
$$
with $\tau$-independent $s_k^f(h)$ - i.e. the $h$-dependence is fully
separated from $\tau$-dependence.
This property will not be true for the basis of ${\cal T}^{\rm DW}_n$.

\subsubsection{Appendix. Rational Calogero (cusp singularity on the
{\it bare} spectral curve) \label{racal}}

It is instructive to consider the simple ''rational'' limit of Calogero system,
obtained when both periods of $E(\tau)$  become large:
$w_1, w_2 \longrightarrow \infty$. Then
$$
F(q|\xi) \longrightarrow \left(\frac{1}{z} - \frac{1}{q}\right)e^{q/z},
$$
$$
\sigma(\xi) \longrightarrow z, \ \ \
\zeta(\xi) \longrightarrow   \frac{1}{z}, \ \ \
\wp(\xi) \longrightarrow \frac{1}{z^2}, \ \ \
\wp'(\xi) \longrightarrow -\frac{2}{z^3},
$$
and the curve $E(\tau)$ degenerates into $y^2 = x^3$.

Remarkably, the polinomials ${\cal T}_n(t)$ remain non-trivial in this limit:
since
$$
\partial^{r_s - 2}\wp(\xi) \ \longrightarrow\
(-)^{r_s}\frac{(r_s-1)!}{z^{r_s}},
$$
eq.(\ref{explT}) gives
\be
t^{-n}{\cal T}_n^{\rm rat}(t|z) = 1 \ +\
\sum_{\stackrel{2\leq r_1 < r_2 < \ldots}{m_s > 0}}
\frac{n!}{ \left(n - \sum_s m_sr_s\right)!}
\prod_s \frac{(-)^m_s}{m_s!(r_s)^{m_s}}
\left(-\frac{g}{zt}\right)^{\sum_s m_sr_s}
\label{explTrat}
\ee
e.g. ${\cal T}^{\rm rat}_0 = 1$,  ${\cal T}^{\rm rat}_1 = t$,
${\cal T}^{\rm rat}_2 = t^2 - \frac{g^2}{z^2}$,
${\cal T}^{\rm rat}_3 = t^3 - \frac{3g^2t}{z^2} + \frac{2g^3}{z^3}$, $\ldots$.

This limit can be also used to illustrate
the derivation of (\ref{2-prod}) and (\ref{r-Sprod}).
Indeed, it is just straightforward to check:
$$
F^{\rm rat}_{12}F^{\rm rat}_{21} = \left(\frac{1}{z} - \frac{1}{q_{12}}\right)
\left(\frac{1}{z} - \frac{1}{q_{21}}\right) =
\frac{1}{z^2} - \frac{1}{q_{12}^2}
$$
- in accordance with (\ref{2-prod}). Further,
\be
F^{\rm rat}_{12}F^{\rm rat}_{23} F^{\rm rat}_{31} =
\left(\frac{1}{z} - \frac{1}{q_{12}}\right)
\left(\frac{1}{z} - \frac{1}{q_{23}}\right)
\left(\frac{1}{z} - \frac{1}{q_{31}}\right) = \nn \\
= \frac{1}{z^3} - \frac{1}{z^2}\left(\frac{1}{q_{12}} +
\frac{1}{q_{23}} +\frac{1}{q_{31}}\right)
- \frac{1}{q_{12}q_{23}q_{31}}.
\nn
\ee
Note that the term with $\frac{1}{z}$ vanishes.
Because of this the symmetric combination
$$
({\cal S}^3F^{\rm rat})_{123} \equiv
F^{\rm rat}_{12}F^{\rm rat}_{23} F^{\rm rat}_{31}  +
F^{\rm rat}_{13}F^{\rm rat}_{32} F^{\rm rat}_{21}  = \frac{2}{z^3}
\longleftarrow \ -\wp'(\xi),
$$
in accordance with (\ref{r-Sprod}).
The idea of the general proof of (\ref{r-Sprod}) can be also illustrated
with this example.
Consider the residue at pole $q_{12} = 0$. In case of
$F^{\rm rat}_{12}F^{\rm rat}_{23} F^{\rm rat}_{31} $ it is equal to
$-\left.F^{\rm rat}_{23} F^{\rm rat}_{31}\right|_{1=2} =
- F^{\rm rat}_{23} F^{\rm rat}_{32}$, while in the case of
$F^{\rm rat}_{13}F^{\rm rat}_{32} F^{\rm rat}_{21} $ it is rather
$+ \left.F^{\rm rat}_{13} F^{\rm rat}_{32}\right|_{1=2} =
+ F^{\rm rat}_{23} F^{\rm rat}_{32}$. Thus this pole disappears
 from $({\cal S}^3F)$, as well as the other would be poles at
$q_{23}=0$ and $q_{13}=0$.
We conclude that  $({\cal S}^3F)$ does not depend of $q$'s.
The reasoning is just the
same for all $({\cal S}^rF)$, as $r>2$ they all are $q$-independent,
and their $z$-dependence is not difficult to work out:
$$
({\cal S}^rF^{\rm rat})(z) = \frac{(r-1)!}{z^r}  \longleftarrow \
\left(-\frac{d}{d\xi}\right)^{r-2}\wp(\xi), \ \ \ {\rm for}\ r > 2.
$$

\subsection{From elliptic Calogero (gauge $N=4$) to Toda-chain
(gauge $N=2$ without matter)
\label{CtoTC}}

Of more interest are asymmetric limits of Calogero system, when the ratio of
periods $\tau = \frac{w_2}{w_1} \rightarrow i\infty$
(see \cite{Ino} and references therein).
Since $w_1$ is finite it is convenient to fix it, the natural choice being
$w_1 = -i\pi$ (thus $w_2 = w_1\tau = -i\pi\tau$). Then
\be
\wp(\xi) = \left(\frac{\pi}{w_1}\right)^2 \left(\sum_{-\infty < M < \infty}
\frac{1}{\sin^2 \frac{\pi}{w_1}(\xi + Mw_2)} - C(\tau)\right)
\stackrel{w_1 = -i\pi}{=}  \nn \\ =
\sum_{-\infty < M < \infty} \frac{1}{\sinh^2(\xi + Mw_2)} + C(\tau) ,
\label{wptri}
\ee
where
\be
C(\tau) = \frac{1}{3} + 2\sum_{M\geq 1} \frac{1}{\sin^2\frac{\pi}{w_1}Mw_2} =
\frac{1}{3} + 2\sum_{M \geq 1} \frac{1}{\sin^2 \pi M\tau} = \nn \\
= \frac{1}{3}\left(1 - 24
\sum_{M\geq 1}\frac{e^{2\pi i M\tau}}{(1 - e^{2\pi i M\tau})^2}\right).
\label{Cfdef}
\ee
If now $\tau \rightarrow i\infty$ ($w_2 \rightarrow +\infty$)
and $\xi$ is kept finite, there is only one term
surviving in the sum: the one with $M=0$:
$\wp(\xi) \rightarrow  (\sinh\xi)^{-2}$.
This gives trigonometric Calogero system with potential
$$
V^{\rm tri}(q) = -g^2\sum_{j<k}^n\frac{1}{\sinh^2 q_{jk}}.
$$

Next, one can consider a scaling limit, when coupling constant $g^2$ grows
along with the growth of $|\tau|$: $g = ig_0 e^{\kappa w_2} = ig_0
e^{-i\kappa\pi\tau}$. In order to keep $V^{\rm tri}$ finite one should require
that {\it all} $\ \ {\rm Re}\ q_{jk} \geq
\kappa w_2 + O(1)$  (since  $(\sinh q)^{-2}
 =  e^{-2q}(1 + O(e^{-2q}))$. This condition
can not be saturated  for all the pairs
$(j,k)$ at once, because $q_{jl} = q_{jk} + q_{kl}$. The only option is to take
\be
q_{jk} = \kappa w_2(k-j) + \frac{1}{2}\hat q_{jk} \ \ \ {\rm for}\  j<k
\label{qhatq}
\ee
(i.e. $q_j = -j\kappa w_2 + \frac{1}{2}
\hat q_j$ with already finite $\hat q_j$).
Then only terms with $|j-k| = 1$ will contribute to $V^{\rm tri}$ in the limit
$w_2 \rightarrow +\infty$:
\be
V_0^{\rm TC}(\hat q) = \lim_{\tau \rightarrow i\infty} V^{\rm tri}(q) =
g_0^2 \lim_{w_2\rightarrow +\infty} \sum_{j<k} \frac{e^{2\kappa w_2}}
{\sinh^2(\kappa w_2(k-j) + \frac{1}{2}\hat q_{jk})} = \nn \\
= g_0^2 \sum_{j=1}^{n-1} e^{-\hat q_{j,j+1}} =
g_0^2 \sum_{j=1}^{n-1} e^{\hat q_{j+1} - \hat q_j}.
\nn
\ee
This is the potential of the Toda-chain, but the chain is not closed:
there is no term $e^{\hat q_1 - \hat q_n}$ in $V_0^{\rm TC}$.

There is, however, a loophole in the above reasoning:
one should better make the substitution (\ref{qhatq}) in the full
Calogero potential, $V^{\rm ell}(q) = -g^2\sum_{j<k}\wp(q_{jk})$,
not in its limit $V^{\rm tri}(q)$. Then the finiteness of the potential
implies that
$|\kappa w_2(k-j) + Mw_2| \geq |\kappa w_2|$ fro all $1\leq j < k \leq n$
and all $M$. This inequality is usually saturated only for
$M=0$, $|j-k| = 1$, but for the single specific value of
scaling index $\kappa = \frac{1}{n}$
an extra opportunity arises: $M=-1$, $k-j = n-1$, i.e. $j=1$, $k=n$.
This provides the lacking term in the potential:
\be
V^{\rm TC}(\hat q) =
\left.\lim_{\tau \rightarrow i\infty} \left(V^{\rm ell}(q)  +
\frac{n(n-1)}{2}g^2\left(\frac{\pi}{iw_1}\right)^2
C(\tau)\right)  \right|_{\stackrel{\kappa = \frac{1}{n}}{w_1 = -i\pi}} =
V_0^{\rm TC}(\hat q) + g_0^2e^{\hat q_1 - \hat q_n}.
\label{VTCfromVell}
\ee

Thus the scaling limit exists, provided $g \sim e^{-i\kappa\pi \tau} =
e^{-\frac{i\pi\tau}{n}}$. Together with the identification (\ref{gverm})
this implies that what is finite as $\tau \rightarrow i\infty$ is the product
$g_0^{2n} \sim g^{2n}e^{2\pi i \tau}  \sim m^{2n}e^{2\pi i \tau}$ -
in  nice agreement with
(\ref{mtscaling}), since $b_1= 2{\rm h}_{SL(n)}^\vee = 2n$.
Note that in order to take the limit in (\ref{VTCfromVell}),  we had to add  a
$q$-independent piece, proportional to $C(\tau)$, to $V^{\rm ell}$ (or, what is
the same, to the Hamiltonian $h_2^{N=4}$).
In other words, the scaling limit involves also redefinition of the moduli:
$h_k^{N=4} $ are not just the same as $h_k^{N=2}$. As soon as the mass $m$
is kept finite, there is a shift:\footnote{
Unfortunately, here  we find some disagreement with \cite{SW2}.
In the case of $SL(2)$ ($n=2$) \cite{SW2} suggests
a formula for $h_2^{N=2} - h_2^{N=4}$
(it is $\frac{1}{2}(u - \tilde u)$ in notations
of \cite{SW2},  eq.(16.25))
which differs from (\ref{2ver4}) by a substitution of $C(\tau)$ by
$$
\frac{1}{2} \hat e_1(\tau) =
\frac{1}{2}\left(\frac{w_1}{\pi}\right)^2 e_1^0(\tau) =
\frac{1}{6}(\theta_{00}^4 + \theta_{01}^4) =
\frac{1}{3} \left(\theta_{00}^4 - \frac{1}{2}\theta_{10}^4\right) =
\frac{1}{3}\left(
1 - 24\sum_{M\geq 1}\frac{(-)^Me^{2\pi i M\tau}}
{(1 - e^{2\pi i M\tau})^2}\right)
= 2C(2\tau) - C(\tau).
$$
Though very similar, this is not our $C(\tau)$ in (\ref{Cfdef}).
The reasoning of \cite{SW2} at this particular point is somewhat
obscure for us: it is not clear,
whether within that approach one can determine
anything but the value of
$\frac{1}{2}e^0_1(\tau = +i\infty) = C(\tau = +i\infty)$.
Note, that in varience with $e^0_1(\tau)$,
our $C(\tau)$ does not have nice modular properties.}
\be
h_2^{N=2} = h_2^{N=4}  + \frac{n(n-1)}{2}\frac{m^2}{8}C(\tau),  \nn \\
h_3^{N=2} = h_2^{N=4} + n(n-1)\frac{m^2}{8}C(\tau), \nn \\
\ldots
\label{2ver4}
\ee

\section{Conclusion}

To conclude, we reviewed the present stage of {\it solid} knowledge at the
''boundary of sciences": between supersymmetric confinement theory
and  integrability theory. Many interesting, but  not yet fully understood
speculations, are not included.
Moreover, we carefully avoided any discussion of the ways, which led to
discovery of Seiberg and Witten, in particular there was no
mentioning of $S$-duality. We did the same in the section,
devoted to integrability theory. Our goal was to extract the {\it results} of
both approaches, and make it clear that they indeed coincide.
The evidence for this coincidence is striking, while its origin is
obscure, and hopefully this should stimulate more people to pay
attention to the role of integrability (and other string theory
implications) in the modern understanding of non-linear quantum dynamics.

As to straightforward developements on the lines of the present paper,
they include examination of Hamiltonian flows, Whitham  flows,
their relation to (geometrical) quantization, the theory of  prepotentials and
''quasiclassical $\tau$-functions'' and - in somewhat orthogonal direction -
generalizations to affine  algebras, which from the
''target space'' perspective should take us from the $4d$ SUSY field theories
to superstrings.  At even simpler level, we mention two kinds of issues,
not yet identified properly in the context of integrability theory: the $N=2$
SUSY models with matter in the fundamental (perhaps, also some gauge
theories with softly broken $N=2$ SUSY), and reinterpretation of the
$N=4\ \longrightarrow \ N=2$ flow in terms of Calabi-Yau manifolds \cite{CY}.
There are many folklor ideas about all these subjects, but we
leave uncomplete results beyond the scope of these notes.

\section{Acknowledgements}

We are indebted for discussions to
O.Aharony,  E.Akhmedov, L.Alvarez-\-Gaume, S.Das, A.Gera\-simov, C.Gomez,
A.Gorsky, A.Hanany, I.Kri\-che\-ver, H.Kunitomo, A.Mar\-shakov, A.Mironov,
K.Oh\-ta, M.Ol\-sha\-netsky, A.Rosly, V.Rubtsov, J.Sonnenschein, A.Stoyanovsky
and A.Tokura.

H.I. is supported in part by Grant-in-Aid for Scientific Research
(07640403) from the Ministry of Education, Science and Culture, Japan.
A.M. acknowledges the
hospitality of  Osaka University and support of the JSPS.


\begin{thebibliography}{12}

\bibitem{UFN} See, for example, \\
   A.Morozov,  {\it String Theory, what is it}, Rus.Physics Uspekhi
     {\bf 35} (1992) 671 (v.{\bf 62},
     {\it No.8}, p.84-175 of Russian edition); \\
   A.Morozov,  {\it Integrability and Matrix Models}, ibid.
     {\bf  164} (1994) {\it No.1}, 3-62 (Rus.ed.), hep-th/9303139.
\bibitem{SW} N.Seiberg and E.Witten,
     {\it Electric-Magnetic Duality, Monopole Condensation and Confinement in
     $N=2$ Supersymmetric Yang-Mills Theory},
     Nucl.Phys. {\bf B426} (1994) 19-52;
     {\it Err.}: ibid. {\bf B430} (1994) 485-486,  hep-th/9407087.
\bibitem{SW2} N.Seiberg and E.Witten,
     {\it  Monopoles, Duality and Chiral Symmetry Breaking
     in $N=2$ Supersymmetric QCD}, ibid. {\bf  B431} (1994) 484-550,
     hep-th/9408099.
\bibitem{Go} A.Gorsky, I.Krichever, A.Marshakov, A.Mironov et al.,
     {\it Integrability and Exact Seiberg-Witten Solution},
     Phys.Lett. {\bf 355B} (1995) 466-474, hep-th/9505035.
\bibitem{MW1} E.Martinec and N.Warner, {\it Integrable Systems
     and Supersymmetric Gauge Theories}, hep-th/9509161.
\bibitem{NT} T.Nakatsu and K.Takasaki, {\it Whitham-Toda Hierarchy and $N=2$
     Supersymmetric Yang-Mills Theory}, hep-th/9509162.
\bibitem{ET} T.Eguchi  and S.K.Yang, {\it  Prepotentials of $N=2$
     Supersymmetric
     Gauge Theories and Soliton Equations}, hep-th/9510183.
\bibitem{DW} R.Donagi and E.Witten, {\it Supersymmetric Yang-Mills
     Theory and Integrable Systems}, IASSNS-HEP-95-78, hep-th/9510101.
\bibitem{M} E.Martinec, {\it Integrable Structures in Supersymmetric
     Gauge and String Theory}, hep-th/9510204.
\bibitem{GoM} A.Gorsky and A.Marshakov, {\it Towards Effective
     Topological Gauge Theories on Spectral Curves}, ITEP/TH-9/95,
     hep-th/9510224.
\bibitem{MW2} E.Martinec and N.Warner, {\it Integrability in $N=2$ Gauge
     Theory: A Proof}, hep-th/9511052.
\bibitem{KLTY}  A.Klemm, W.Lerche, S.Theisen and S.Yankielowicz,
     {\it Simple Singularities and $N=2$ Supersymmetric Yang-Mills Theory},
     Phys.Lett. {\bf 344B} (1995) 169, hep-th/9411048.
\bibitem{AF} P.Agyres and A.Faraggi, {\it The Vacuum Structure and Spectrum of
     $N=2$ Supersymmetric $SU(N)$ Gauge Theory},
     Phys.Rev.Lett. {\bf 73} (1995) 3931, hep-th/9411057.
\bibitem{DS} U.H.Danielsson and B.Sundborg, {\it The Moduli Space and
     Monodromis of $N=2$ Supersymmetric $SO(2R+1)$ Yang-Mills Theory},
     hep-th/9504102.
\bibitem{BL} A.Brandhuber and K.Landsteiner, {\it On the Monodromies of $N=2$
     Supersymmetric Yang-Mills Theory with Gauge Group $SO(2n)$},
     hep-th/9507008.
\bibitem{HO}  A.Hanany and Y.Oz , {\it  On the Quantum Moduli Space of Vacua
     of $N=2$ Supersymmetric $SU(N_c)$ Gauge Theories}, hep-th/9505075.
\bibitem{OW} D.Olive and E.Witten, {\it Supersymmetry Algebra that Includes
     Topological Charges}, Phys.Lett. {\bf 78B} (1978) 97-101.
\bibitem{OP}  The classical review paper on
     integrable systems of particles is: \\
   M.Olshanetsky and A.Perelomov, {\it
     Classical Integrable Finite-\-Dimensional
     Systems Related to Lie Algebras}, Phys.Rep. {\bf 71C} (1981) 97-101.
\bibitem{Kr} The relevant chapter of integrability theory is very old
     and the number of references is very large.
     We mention just a few papers, directly related to our consideration: \\
   I.Krichever, {\it Methods of
     Algebraic Geometry in the Theory of Nonlinear Equations},
     Sov.Math.Surveys, {\bf 32} (1977) 185-213; \\
   I.Krichever, {\it The Integration of
     Non-linear Equations by the Methods of Algebraic
     Geometry}, Funk.Anal. and Appl.
     {\bf 11} (1977) {\it No.1} 15-31 (Rus.ed.);\\
   I.Krichever, {\it  Elliptic Solutions of the Kadomtsev-Petviashvili Equation
     and Integrable System of Particles}, Funk.Anal. and Appl. {\bf 14} (1980)
     282-290 ({\it No.4} 15-31 of Rus.ed.);\\
   B.Dubrovin, {\it Theta Functions
     and Non-linear Equations}, Sov.Math.Surveys,
     {\bf 36} (1981) {\it No.2} 11-80 (Rus.ed.); \\
   H.Flaschka  and D.McLaughlin, {\it Canonicaly Conjugate Variables for the
     KdV Equation and the Toda Lattice with Periodic Boundary Conditions},
     Progr.Theor.Phys. {\bf 55} (1976) 438-456;\\
   M.Adler and P.van Moerbeke, {\it Completely
     Integrable Systems, Euclidean Lie
     Algebras and Curves}, Adv.Math. {\bf 38} (1980) 267-317; \\
   M.Adler and P.van Moerbeke, {\it Linearization of Hamiltonian Systems,
Jacobi
     Varieties and Representation Theory}, ibid. 318-379. \\
   O.Babelon, E.Billey, I.Krichever and M.Talon,
     {\it Spin Generalization of the
     Calogero-\-Moser System and the Matrix KP Equation}, hep-th/9411160.
\bibitem{Hi} The modern-style language in this field uses the
     notion of Hitchin systems, which empasises interpretation of Lax equation
     as a flat connection condition. A few directly relevant references are:\\
   N.Hitchin, {\it Stable Bundles and Integrable Systems},
     Duke Math.Journ. {\bf 54}
     (1987) 91-114; \\
   N.Hitchin, {\it Flat Connections and Geometric Quantization},
Comm.Math.Phys.
     {\bf 131}  (1990) 347-380; \\
   E.Markman, {\it Spectral Curves and Integrable Systems}, Comp.Math. {\bf 93}
     (1994) 255-290; \\
   A.Beilinson and V.Drinfeld,
     {\it Quantization of Hitchin's Fibration and Langlands
     Program}, preprint (1994);\\
   B.Feigin and E.Frenkel,
     {\it Affine Kac-Moody Algebras at the Critical Level and
     Gelfand-Dikii Algebras},
     Int.J.Mod.Phys. {\bf  A7}, Suppl.{\bf 1A} (1992) 197-215; \\
   B.Enriquez and V.Roubtsov,
     {\it Hitchin Systems, Higher  Gaudin Operators and
     R-Matrices}, alg-geom/9503010;  \\
   N.Nekrasov, {\it Holomorphic Bundles and Many-Body Systems},
     hep-th/9503157; \\
   M.Olshanetsky {\it  Generalized Hitchin Systems and the
     Knizhnik-\-Zamolod\-chikov-\-Bernard Equation
     on Elliptic Curves}, hep-th/9510143; \\
   O.Sheinman, {} hep-th/9510165; \\
     see also \cite{DW} and \cite{M}.
\bibitem{Wh} See the following papers for general description and
     references:\\
   B.Dubrovin and S.Novikov,
     Sov.Math. Surveys {\bf 36} (1981) {\it No.2} 12 (Rus.ed.); \\
     ref.\cite{Kr2}  below.
\bibitem{D} B.Dubrovin, {\it Geometry of $2d$ Topological Field Theories},
     SISSA-89/94/FM;\\
   B.Dubrovin, {\it  Integrable Systems in Topological Field Theory},
Nucl.Phys.
     {\bf B379} (1992) 627-689; \\
   B.Dubrovin, {\it Hamiltonian Formalism of
     Witham-type Hierarchies in Topological
     Landau-Ginzburg Model}, Comm.Math.Phys. {\bf 145} (1992) 195.
\bibitem{Kr2}  I.Krichever, {\it The $\tau$-function of the
     Universal Whitham Hierarchy,
     Matrix Models and Topological Field Theories}, hep-th/9205110;\\
   I.Krichever, {\it The Dispersionless Lax
     Equations in Topological Minimal Models},
     Comm. Math.Phys. {\bf 143} (1992) 415-429.
\bibitem{LP} A.Losev and I.Polyubin, {} JETP Lett. {\bf 58} (1993) 573,
     hep-th/9305079.
\bibitem{KMMM/MPLA} S.Kharchev, A.Marshakov, A.Mironov et al.,
     {\it Landau-Ginzburg Topological Theories in the Framework of
     GKM (Generalized Kontsevich Model) and Equivalent Hierarchies},
     Mod.Phys.Lett. {\bf A8} (1993) 1047,        hep-th/9208046; \\
   S.Kharchev and A.Marshakov, Int.J.Mod.Phys. {\bf A10} (1995) 1219.
\bibitem{STY} J.Sonnenschein, S.Theisen and S.Yankielowicz,
     {\it On the Relation between the Holomorphic Prepotentials and the Quantum
     Moduli in SUSY Gauge Theories}, hep-th/9510129.
\bibitem{Pic} A Ceresole, R.D'Auria and S.Ferrara,
     Phys.Lett. {\bf 339B} (1995) 71; \\
   A.Klemm, W.Lerche and S.Theisen, {\it Nonperturbative Effective Actions
     of $N=2$ Supersymmetric Gauge Theories}, hep-th/9505150;
\bibitem{AS}  See, for example: \\
   M.Abramowitz and I.Stegun, {\it Handbook on Mathematical Functions},
     Dover Publ., NY, 1965;  sect.18.
\bibitem{Ino} V.Inosemtsev, {\it The Finite Toda Lattices},
     Comm.Math.Phys. {\bf 121} (1989) 629-638.
\bibitem{CY} See, for example: \\
   A.Strominger, {\it Massless Black Holes and Conifolds in String Theory},
     hep/th 9504090; \\
   S.Kachru, A.Klemm, W.Lerche, P.Mayr and C.Vafa, {\it Nonperturbative
     Results on the Point-Particle Limit of $N=2$
     Heterotic String Compactification},
     hep/th 9508155; \\
   and ref.\cite{MW2}.


\end{thebibliography}
\end{document}